\documentclass{article}
\usepackage[utf8]{inputenc}
\usepackage[english]{babel}
\usepackage[letterpaper,top=2cm,bottom=2cm,left=3cm,right=3cm,marginparwidth=1.75cm]{geometry}
\usepackage{amsmath}
\usepackage{amsfonts}
\usepackage{amssymb}
\usepackage{graphicx}
\usepackage{authblk}
\usepackage[bookmarks=false]{hyperref}
\usepackage[numbers,sort&compress]{natbib}
\usepackage{enumitem}
\usepackage{subcaption}
\usepackage{tabularray}
\usepackage{multicol}
\usepackage[usenames,dvipsnames]{xcolor}
\usepackage{lipsum}
\definecolor{deep_purple}{HTML}{4702ed}

\newcommand{\changescolor}{deep_purple}

\newcommand{\change}[1]{{\color{\changescolor}#1}}

\newenvironment{changed}{\color{\changescolor}}{}

\graphicspath{{figs}}

\newcommand{\imag}{\mathcal{I}m\;} 
\newcommand{\real}{\mathcal{R}e\;} 
\newcommand{\fref}[1]{Fig.~\ref{#1}}
\newcommand{\tref}[1]{Tab.~\ref{#1}}

\title{Simulation of single diffraction dissociation in resonance region at LHC energies}

\author[1]{O. S. Potiienko}
\author[2]{D. V. Zhuravel}
\author[3]{D. M. Riabov}
\author[1]{N. O. Chudak}

\affil[1]{Odesa Polytechnic National University, Shevchenko av. 1, Odesa, 65044, Ukraine}
\affil[2]{Bogolyubov Institute for Theoretical Physics (BITP), National Academy of Sciences of Ukraine, Metrolohichna st. 14-b, Kiev, 03143, Ukraine}
\affil[3]{State University of Intellectual Technologies and Communications, Kuznechna st. 1, Odesa, 65023, Ukraine}

\begin{document}
\maketitle

\begin{abstract}
A comprehensive review of the single diffraction dissociation duality-based model at low missing masses have been presented.
The distinguishing feature of the model is the nonlinear Regge proton trajectory used to account for the resonances contributions to the cross-sections.
It helps classify and understand the spectrum of excited states of proton and their decays, providing insights into the internal structure and dynamics of particles.
The behavior of the differential cross-section in the resonance region at small missing masses $M_x$ is investiaged.
The model parameters are refined in the light of new experimental data.
\end{abstract}

\section{Introduction}

As is known from many experimental results \cite{Goulianos:1982vk}, in the processes of hadron scattering at high energies, the majority of events are concentrated in the region of small momentum transfers. The combination of high energies and small momentum transfers creates conditions for the realization of diffraction processes \cite{feinberg1956nuovo, Good:1960ba}. Diffraction in particle physics draws an analogy from classical wave diffraction, where waves encounter an obstacle and spread out. Similarly, in diffractive dissociation, a hadron such as a proton encounters another particle or nucleus, leading to its partial or complete dissociation. The process is characterized by a small momentum transfer between the incident particle and the target, resulting in a "rapidity gap" -- a region in the detector with very few or no particles. One of the significant findings from diffractive dissociation studies is the confirmation of the pomeron exchange model, which has been supported by experimental data showing characteristic rapidity gaps in diffractive events. The detailed measurements of diffractive cross-sections and the kinematic distributions of final-state particles have provided insights into the non-perturbative aspects of QCD.

\begin{figure}[!ht]
\begin{center}
\includegraphics[width=0.98\linewidth]{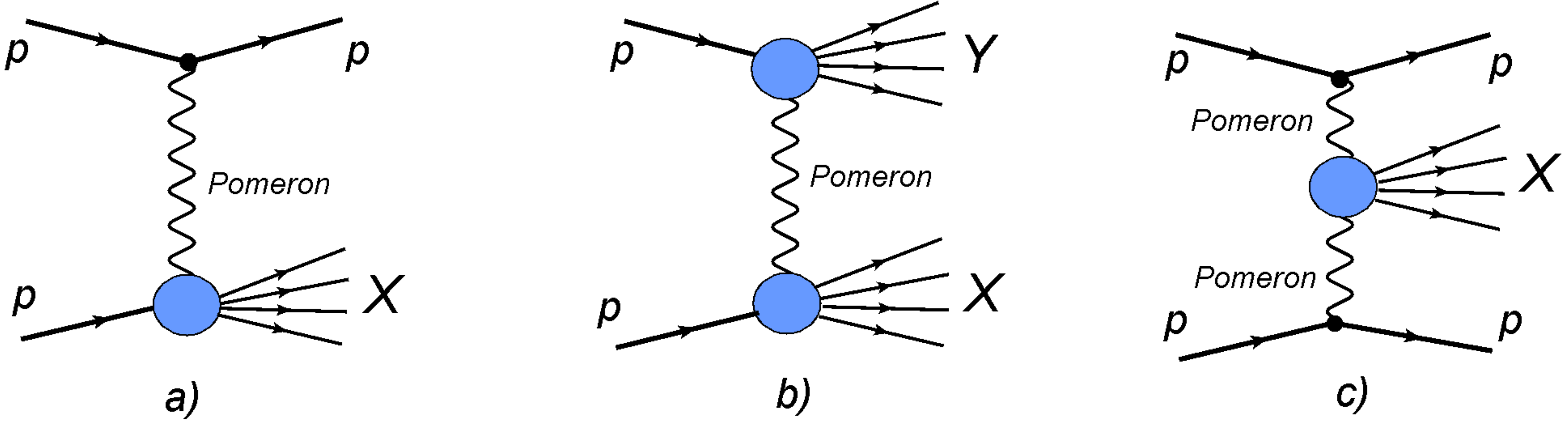}
\caption{The main types of diffractive processes: a) single dissociation (SD), b) double dissociation (DD) and c) central diffraction (CD), where p - is proton, X(Y) - is a system of secondary hadrons.}
\label{fig:diffrac}
\end{center}
\end{figure}

The main types of diffractive processes are single dissociation (SD), double dissociation (DD), and central diffraction (CD) (\fref{fig:diffrac}). Diffraction dissociation has been studied in various high-energy physics experiments, such as TOTEM \cite{Antchev2013}, CMS \cite{cs_khachatryan}, CERN \cite{cs_albrow, cs_ansorge, cs_bernard}, ALICE \cite{cs_abelev} and ATLAS \cite{Aad2020-cw}. In this work, we consider the low-mass single diffraction dissociation of protons. The small missing masses $M_X$ show a multitude of peaks or features corresponding to different nucleon resonances \cite{Jenkovszky2011}. Nucleon resonances are essential for understanding the strong force interactions described by QCD.

The experimentally observed characteristics of the diffraction dissociation largely depend on the properties of the hadron-region vertex. We will consider this vertex as the amplitude of a real process. Using the unitarity condition \cite{BAKER1976}, duality \cite{Collins1977}, this amplitude can be calculated as the sum of the resonance contributions known from experiments \cite{Hagiwara2002}. In this work, we examined the exchange of a single pomeron. This pomeron is typically considered to be a simple pole of the partial scattering amplitude. However, there are other dynamic models, such as the model with exchange of a pomeron-like analog but with an odd signature \cite{Jenkovszky:2018itd, Szanyi:2019kkn}, and the dipole pomeron model, which is the foundation, further enhanced by incorporating a dip-bump mechanism \cite{Jenkovszky:2024avi}. Therefore, it is interesting to explore the combination of different models of the hadron-reggeon vertex and collision dynamics to describe experimental data. In this work, we present the results of fits with the exchange of a single pomeron pole.

Our goal is to calculate the differential cross-section $d\sigma/dt$ of single diffraction dissociation within the resonance region at small missing masses $M_X$ and the total single diffraction dissociation cross section.

The structure of the article is presented as follows. In Section 2 we find the structure function of proton $W_2\left(M_X^2,t\right)$, which related to the imaginary
part of dual-Regge transition amplitude $A\left(M_X^2, t\right)$. In Section 3 we show that the behavior of the cross-section is influenced by baryon resonances. In Section 4 we calculate total and differential cross-sections and fit them to the experimental data. 
\section{From elastic scattering to single diffractive dissociation}

Let us consider the diffractive dissociation $p + p \to p + X$, where $p$ is a proton, $X$ is a system of secondary hadrons. This process is represented as a similar elastic process, where $X$ is considered as a single particle with a squared mass $M_X^2$, which equals the scalar square of the sum of the four-momenta of $X$. Then, instead of the elastic scattering amplitude we obtain the amplitude of diffractive dissociation with the vertices of the reggeon-hadron interaction. Such modification \cite{Jenkovszky2011} leads to the expression \eqref{sddcs} for the differential cross-section of single diffraction dissociation at large $s$ (the square of the center-of-mass energy of the collision)

%
\begin{equation}\label{sddcs}
	\frac{d{{\sigma }_{SD}}}{dtd{{M}_{x}^2}}\approx\frac{9{{\beta }^{4}}}{4\pi }{{\left[ {{F}^{p}}\left( t \right) \right]}^{2}}{{\left( \frac{s}{M_{X}^{2}} \right)}^{2\left( {{\alpha }_{P}}\left( t \right)-1 \right)}}\frac{{{W}_{2}}\left( t,{{M}_{x}^2} \right)}{2m}\,,
\end{equation}
where $t$ is the momentum transfer between colliding particles, $\beta $ is the quark-Pomeron coupling, $m$ is proton mass, 
${{\alpha }_{P}}\left( t \right)$ is a vacuum Regge trajactory, ${{F}^{p}}\left( t \right)$ is elastic form factor of proton, ${{W}_{2}}\left( t,{{M}_{x}} \right)$ is a structure function of proton, which describes the Pomeron-proton vertex. 
According to \cite{Jenkovszky2011, Donnachie2002} we take the Pomeron trajectory ${{\alpha }_{P}}\left( t \right) = 1.08 + 0.25t$.
The expression for the proton elastic form factor is given by ${{F}^{p}}\left( t \right) = \left( 1.0 - t/0.71 \right)^{-2}$.

\begin{figure}[!ht]
\begin{center}
\includegraphics[width=0.4\linewidth]{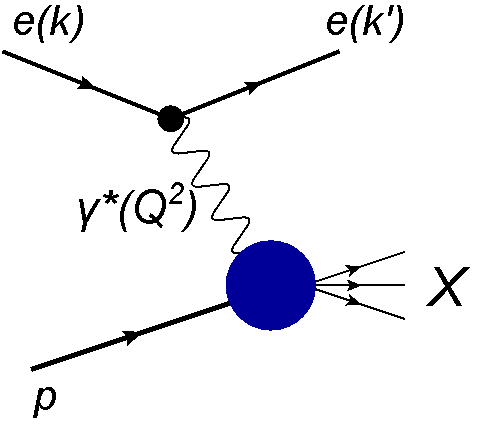}
\caption{Deeply virtual Compton scattering process $\gamma^*p \to X$ transition}
\label{fig:PomPhot}
\end{center}
\end{figure}

The most challenging building block of the expression \eqref{sddcs} is the structure function ${{W}_{2}}\left( t,{{M}_{x}^2} \right)$.

Following \cite{Jenkovszky2011}, this structure function can be constructed from the similarity between the Pomeron-proton diffractive proccess $\mathbb{P}p \to X$ (Pomeron denoted by $\mathbb{P}$) and deeply virtual Compton scattering process $\gamma^*p \to X$ (\fref{fig:PomPhot}). 
Let us briefly highlight the major steps on the way to $W_2\left(M_X^2,t\right)$:

\subsubsection*{\textit{Step 1: Similarity with $\gamma^*p \to X$}}

We utilize the similarity between the inelastic vertices $\mathbb{P}p \to X$ and $\gamma^*p \to X$ and the corresponding structure functions.
The structure functions $F_2\left(x, Q^2\right)$ for $\gamma^*p$ processes has been studied in \cite{Fiore_2004}, 
and are connected \cite{Halzen1984} with $W_2$ through the following relation
\begin{equation}\label{W2F2relation}
\nu W_2\left(M_X^2,t\right) = F_2\left(x, t\right),
\end{equation}
\noindent
where $\nu = \left(M_X^2 - m^2 - t\right)/2m$ is another kinematic variable. 
Thus, establishing an analogy between the virtual photon $\gamma^*$ and the Pomeron ($\mathbb{P}$), and setting $Q^2 = -t$, yields
\begin{equation}\label{w21}
\nu W_2\left(M_X^2,t\right) = F_2\left(x, t\right) = 
\frac
{-4t\left(1-x\right)^2}
{\alpha_{fs} \left(M_X^2 - m^2\right)\left(1 - 4m^2x^2/t\right)^{3/2}}
\imag A\left(M_X^2, t\right),
\end{equation}
\noindent
where $\alpha_{fs}$ is fine structure constant, $x = -t / 2m\nu$ is Bjorken variable, and $A\left(M_X^2, t\right)$ is the dual-Regge transition amplitude.

\subsubsection*{\textit{Step 2: Dual-Regge amplitude}}

The amplitude $A\left(M_X^2, t\right)$ in eq. \eqref{w21} is obtained from unitarity and Veneziano duality \cite{Jenkovszky2011}.
According to the model of dual amplitude with Mandelstam analiticity (DAMA), the amplitude is expressed as the sum over Regge trajectories, ending up with the direct channel resonance decomposition
\begin{equation}\label{A1}
A\left(M_X^2,t\right) = a \sum_{n \geqslant 0}
\frac
{\left[f(t)\right]^{2(n + 1)}}
{2n + 0.5 - \alpha\left(M_X^2\right)},
\end{equation}
\noindent
where $a$ is normalization factor, $f(t) = (1 - t/t_0)^{-2}$ is the form factor of $\mathbb{P}p \to \mathbb{P}p$ system, $t_0$ is the model parameter. 
Finally, $\alpha\left(M_X^2\right)$ is the (nonlinear complex) baryonic resonance trajectory in $M_X^2$ channel, which allows to account for a set of various resonances.

The imaginary part of the amplitude \eqref{A1} needed for structure function \eqref{w21} is given by
\begin{equation}\label{A1im}
\imag A\left(M_X^2,t\right) = 
a \sum_{n \geqslant 0}
\left[f(t)\right]^{2(n + 1)}
\frac
{\imag \alpha\left(M_X^2\right) }
{\left[2n + 0.5 - \real \alpha\left(M_X^2\right)\right]^2 + \left[\imag \alpha\left(M_X^2\right)\right]^2}.
\end{equation}


\subsubsection*{\textit{Step 3: Proton trajectory}}

\begin{figure}[!ht]
\centering
\begin{subfigure}{.48\textwidth}
  \centering
  \includegraphics[width=\linewidth]{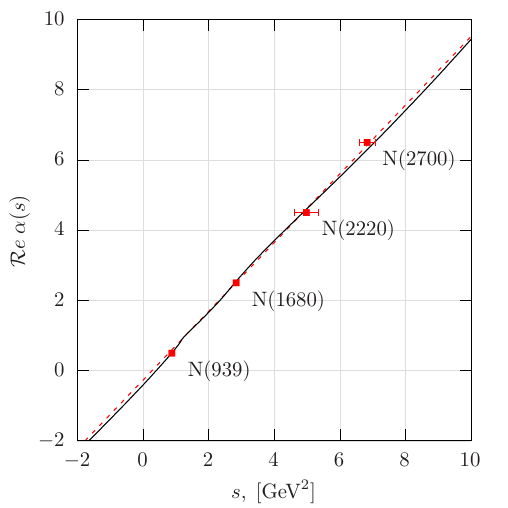}
  \caption{The real part of fitted proton trajectory \eqref{rea} (black solid line) and linear fit (red dashed line).}
  \label{fig:real_alpha}
\end{subfigure}%
\hfill
\begin{subfigure}{.48\textwidth}
  \centering
  \includegraphics[width=\linewidth]{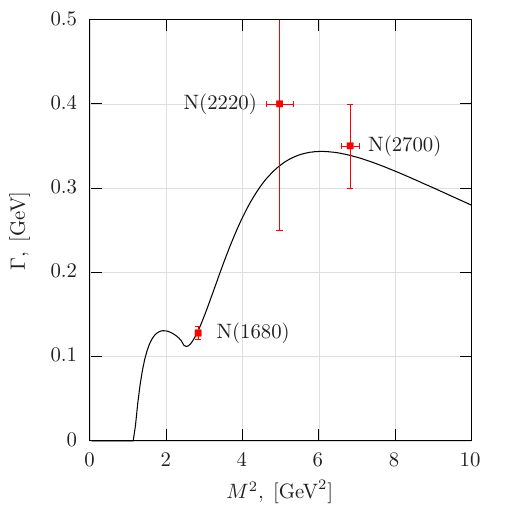}
  \caption{The fit of resonance widths \eqref{width} (solid line) and experimental data from \tref{tab:baryons}.}
  \label{fig:widths}
\end{subfigure}
\caption{\centering The fit \cite{Fiore_2004_trajectory} of the proton trajectory to N-resonances data. }
\label{fig:trajectory}
\end{figure}

The last step in this journey is the nonlinear complex proton Regge trajectory $\alpha\left(M_X^2\right)$.
It helps classify and understand the spectrum of excited states of proton and their decays, providing insights into the internal structure and dynamics of particles.
These resonances indicate specific energy and angular momentum states where the proton can undergo transitions to excited states before returning to the ground state.
The real part of the trajectory $\real \alpha\left(M_X^2\right)$ provides the relation between the mass of resonance and its angular momentum (quantum number $J$). At the same time, the imaginary part of the trajectory provides the Breit-Wigner widths of the resonances 
\begin{equation}\label{width}
\Gamma = \frac{\imag \alpha\left(M^2\right)}{M \real \alpha'\left(M^2\right)},
\end{equation}
\noindent
where $M$ is the resonance mass, $\real \alpha'\left(M^2\right)$ denotes the first derivative (the slope) of the real part of trajectory.
The problem is to find a trajectory $\alpha\left(M^2\right)$ featuring both the almost linear real part (\fref{fig:real_alpha}) and essentialy nonlinear imaginary part (\fref{fig:widths}), while holding the analiticity properties.
Such trajectory has been extensively studied in \cite{Fiore_2004_trajectory} with the help of dispersion relations, which provides us with the expressions
\begin{equation}\label{ima}
\imag \alpha\left(s\right) = 
s^\delta
\sum_{n}c_n
\left(\frac{s - s_n}{s_n}\right)^{\real \alpha\left(s_n\right)}
\theta\left(s - s_n\right),
\end{equation}

\begin{equation}\label{rea}
\real \alpha\left(s\right) = 
\alpha(0) + \frac{s}{\pi}
\sum_n c_n \mathcal{A}_n(s),
\end{equation}

\noindent
where $c_n$ and $s_n$ are parameters to be fitted using the experimental data on resonances widths and masses,
$\alpha(0)$ is the intercept of the real part of trajectory.
Note, that hereinafter the quantities $s$ and $s_n$ are nondimentionalized by $s_0 = 1\;\text{GeV}$, so that resulting expressions are dimensionless, e.g. \eqref{ima} and \eqref{rea}.
The Heaviside step function $\theta(\cdot)$ is conventionally defined such that $\theta\left(0\right) = \frac{1}{2}$. 
The term $ \mathcal{A}_n(s)$ in \eqref{rea} emerges from a dispersion relation and is given by
\begin{eqnarray} \label{An}
\begin{aligned} 
\mathcal{A}_n(s) = 
&
\frac{\Gamma(1 - \delta)\Gamma(\lambda_n + 1)}{\Gamma(\lambda_n - \delta + 2)}
s_n^{\delta - 1}
{{}_2F_1\left(1, 1-\delta; \lambda_n - \delta + 2; \frac{s}{s_n}\right)}
\theta\left(s_n - s\right)
+ \\
&
\left\{
\pi s^{\delta - 1}\left(\frac{s - s_n}{s}\right)^{\lambda_n} \cot \left[\pi\left(1 - \delta\right)\right] - \right. \\
&\left.
\frac{\Gamma(-\delta)\Gamma(\lambda_n + 1)}{\Gamma(\lambda_n - \delta + 1)}
\left(\frac{s_n^{\delta}}{s}\right)
{}_2F_1\left(\delta - \lambda_n, 1; \delta + 1; \frac{s_n}{s}\right)
\right\}
\theta\left(s - s_n\right),
\end{aligned}
\end{eqnarray}
\noindent
where $\delta$ is the dimensionless paramter to be fitted, $\Gamma\left(x\right)$ is the gamma function, ${}_2F_1\left(a, b; c, z\right)$ is the Gaussian hypergeometric function, $\cot\left(x\right)$ denotes cotangent function, and $\lambda_n = \real \alpha \left(s_n\right)$.

Note that $\real \alpha$ appear on the both sides of equation \eqref{rea} through $\lambda_n$.
Thus, \eqref{ima}, \eqref{rea} and \eqref{An} are in fact functional equations, which significantly sophisticates the fitting procedure necessary to determine the value of parameters $\alpha(0)$, $\delta$, $c_n$, and $s_n$, where $n = 1, 2, x$. 

The baryons included in the trajectory are N(939), N(1680), N(2220), N(2700) and their properties depicted in \tref{tab:baryons}. 
These resonances have been observed in experiments studying single diffraction dissociation processes, providing a framework for the study of SDD.
The recursive fitting procedure of proton trajectory \eqref{ima}, \eqref{rea} to the data for these resonances has been performed in \cite{Fiore_2004_trajectory}. 

\begin{table}[!ht]
\centering
\begin{tblr}{colspec={Q[c, 2cm] Q[c, 2cm] Q[c, 3cm] Q[c, 3cm]}, rowsep = 4pt,}
\hline
Name & $J$ & $M$ (MeV) & $\Gamma$ (MeV) \\
\hline 
N(939) & $1/2$ & $939$ &  $-$ \\
N(1680) & ${5}/{2}$ & $1684 \pm 4$ & $128 \pm 8$ \\
N(2220) & ${9}/{2}$ & $2230 \pm 80$ & $400 \pm 150$ \\
N(2700) & ${13}/{2}$ & $2612 \pm 45$ & $350 \pm 50$ \\
\hline
\end{tblr}
\caption{The baryons included in the fit \cite{Fiore_2004_trajectory} of trajectory $\alpha(s)$. The columns from left to right are: the baryon name, total angular momentum $J$, Breie-Wigner mass $M$, Breit-Wigner width $\Gamma$.}
\label{tab:baryons}
\end{table}

\textit{The fitting algorithm} can be summarized as follows.
Due to the close-to-linear form of the real part of trajectory, it is reasonable to start with the fit of the linear approximation of the real part $\alpha_{lin}(s) = \alpha(0) + \alpha'(0)s$, which gives the initial values of $\lambda^{(0)}_n = \real \alpha_{lin}\left(s_n\right)$.
Then, the expression \eqref{width} for the widths of resonances is iteratively fitted to the experimental data \tref{tab:baryons}.
After each iteration, the values of $\lambda_n$ are updated $\lambda_n = \real \alpha \left(s_n\right)$ using \eqref{rea} with the updated values of parameters. 
The process is repeated until the sequence of trajectories converges.
The resuling values of parameters are summarized in below in \tref{tab:trajectory_parameters}.

\begin{table}[!ht]
\centering
\begin{tblr}{colspec={Q[l, 2.5cm] Q[l, 2.5cm] Q[l, 2.5cm]}, rowsep = 2pt}
\hline 
$\alpha(0) = -0.41$ 			& $c_1 = 0.51$ 					& $s_1 = 1.16\;\text{GeV}^2$  \\
$\;\;\;\;\;\delta = -0.46$     	& $c_2 = 4.0$ 					& $s_2 = 2.44\;\text{GeV}^2$  \\
								& $c_x =  4.6 \times 10^3$ 		& $s_x = 11.7\;\text{GeV}^2$  \\
\hline
\end{tblr}
\caption{Parameters of proton trajectory $\alpha(s)$ obtained in \cite{Fiore_2004_trajectory} by fitting \eqref{width} to resonances data \tref{tab:baryons}.}
\label{tab:trajectory_parameters}
\end{table}

Using the values of parameters from \tref{tab:trajectory_parameters} we plot the trajectory $\alpha(s)$ and ensure it match \cite{Fiore_2004_trajectory} (see \fref{fig:trajectory}). 
This completes the construction of the expression for differential cross-section for single diffraction dissociation \eqref{sddcs}.

Substituting $W_2$ from \eqref{w21} into \eqref{sddcs} we obtain

\begin{eqnarray}\label{diffcs}
\begin{aligned} 
\frac{d^2\sigma}{dtdM_X^2}\left(M_X^2, t\right)=
A_0 
\left(\frac{s}{M_X^2}\right)^{2\alpha_P(t) - 2}
\frac{x\left(1 - x\right)^2\left[F^p(t)\right]^2}{\left(M_X^2 - m^2\right)\left(1 - 4m^2x^2/t\right)^{3/2}}&
\\
\times \sum_{n=1}^{3}
\left[f(t)\right]^{2(n + 1)}
\frac
{\imag \alpha\left(M_X^2\right) }
{\left[2n + 0.5 - \real \alpha\left(M_X^2\right)\right]^2 + \left[\imag \alpha\left(M_X^2\right)\right]^2}&,
\end{aligned} 
\end{eqnarray}

\noindent
where $A_0 = 9a\beta^4/\pi\alpha_{fs}$ is the normalization factor that combines the factors $a$ and $\beta$ from \eqref{sddcs} and \eqref{A1im}.
Another free parameter is $t_0$ of the form-factor $f(t)$ appearing in the expression for the transition amplitude \eqref{A1}.
The rest of parameters $\alpha(0), \delta, c_n, s_n$ ($n = 1, 2, x$) are fixed from trajectory fit \tref{tab:trajectory_parameters}.
The values of $A_0$ and $t_0$ can be determined from the fit of the differential cross-section $d\sigma_{\text{SDD}}/dt$ and total cross-section $\sigma_{\text{SDD}}$.
In the next section we analyze the behavior of the double differential cross-section \eqref{diffcs} and discuss the applicability range of the model.
\section{Model behavior in the resonance region}
\begin{figure}[!ht]
\begin{center}
\includegraphics[width=0.5\linewidth]{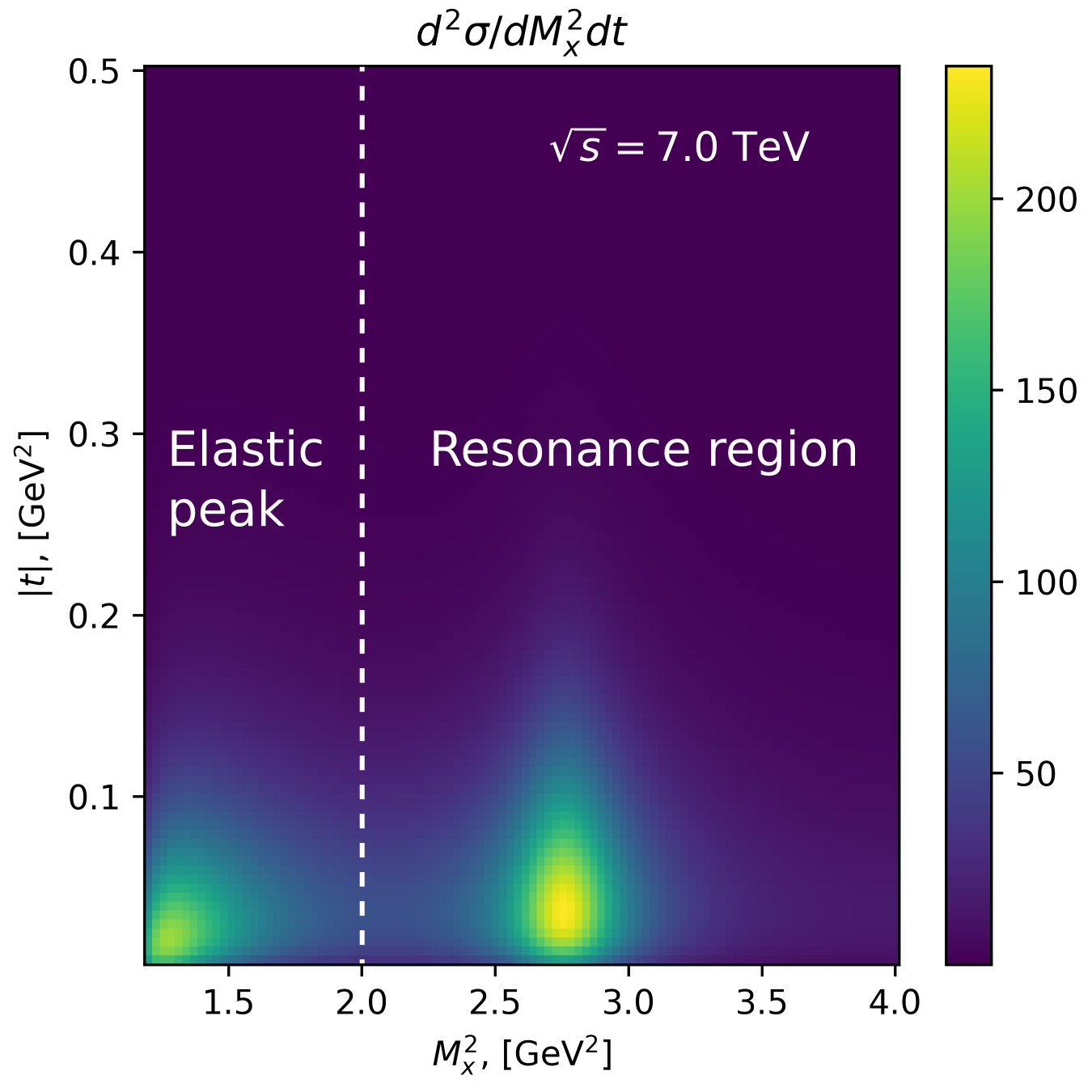}
\caption{
The differential cross-section $d^2\sigma/dtdM_X^2$ \eqref{diffcs} in $t$, $M_X^2$ plane. There are two missing mass $M_X$ regions: elastic peak region ($M_X^2 < 2\;\text{GeV}^2$), and the resonance region ($2\;\text{GeV}^2 M_X^2 \leqslant 8\;\text{GeV}^2$).}
\label{fig:regions}
\end{center}
\end{figure}

In the study of single diffraction dissociation processes the regions of small and large missing masses $M_X$ offer distinct mechanisms of particle interactions. 
Here we focus on the small missing masses $M_X$, where the behavior of the cross-section is notably influenced by resonances within the structure functions.
These resonances manifest as peaks in the cross-section data, shaping the scattering processes observed at lower energies. 
In the present model, the resonances are accounted via proton trajectory described in the previous section.
This trajectory defines the behavior of the differential cross-section \eqref{diffcs} in the $M_X$ region where resonances appear \fref{fig:regions}.
As $M_X \to m$, the scattering process exhibits elastic-like nature ($m$ is the proton mass), which significantly differs from the resonance production processes.
The proper account for the contributions from the elastic and close-to-elastic processes requires a distinct model.
Thus we focus on the contributions from the resonance region only and \textit{consider elastic processes as background contributions}. 
In the present model the elastic contributions appear as the elastic peak in the differential cross-section as low $M_X$, as can be seen in \fref{fig:regions}.
To avoid double counting in the elastic region, we follow \cite{Jenkovszky2011} and consider $M_X^2 \geqslant 2\;\text{GeV}^2$, which cuts elastic peak off.
At the same time, at large missing masses $M_X$, the contributions of resonance productions expectedly reduce, making cross-section \eqref{diffcs} neglectedly small for $M_X^2 > 8\;\text{GeV}^2$.
In this region another Regge mechanisms come into play, but they are outside of the scope of the present work.
This justifies what we call the \textit{resonance region}  $2\;\text{GeV}^2 \leqslant M_X^2 \leqslant 8\;\text{GeV}^2$.

\begin{figure}[!ht]
\centering
\begin{subfigure}{.49\textwidth}
  \centering
  \includegraphics[width=1\linewidth]{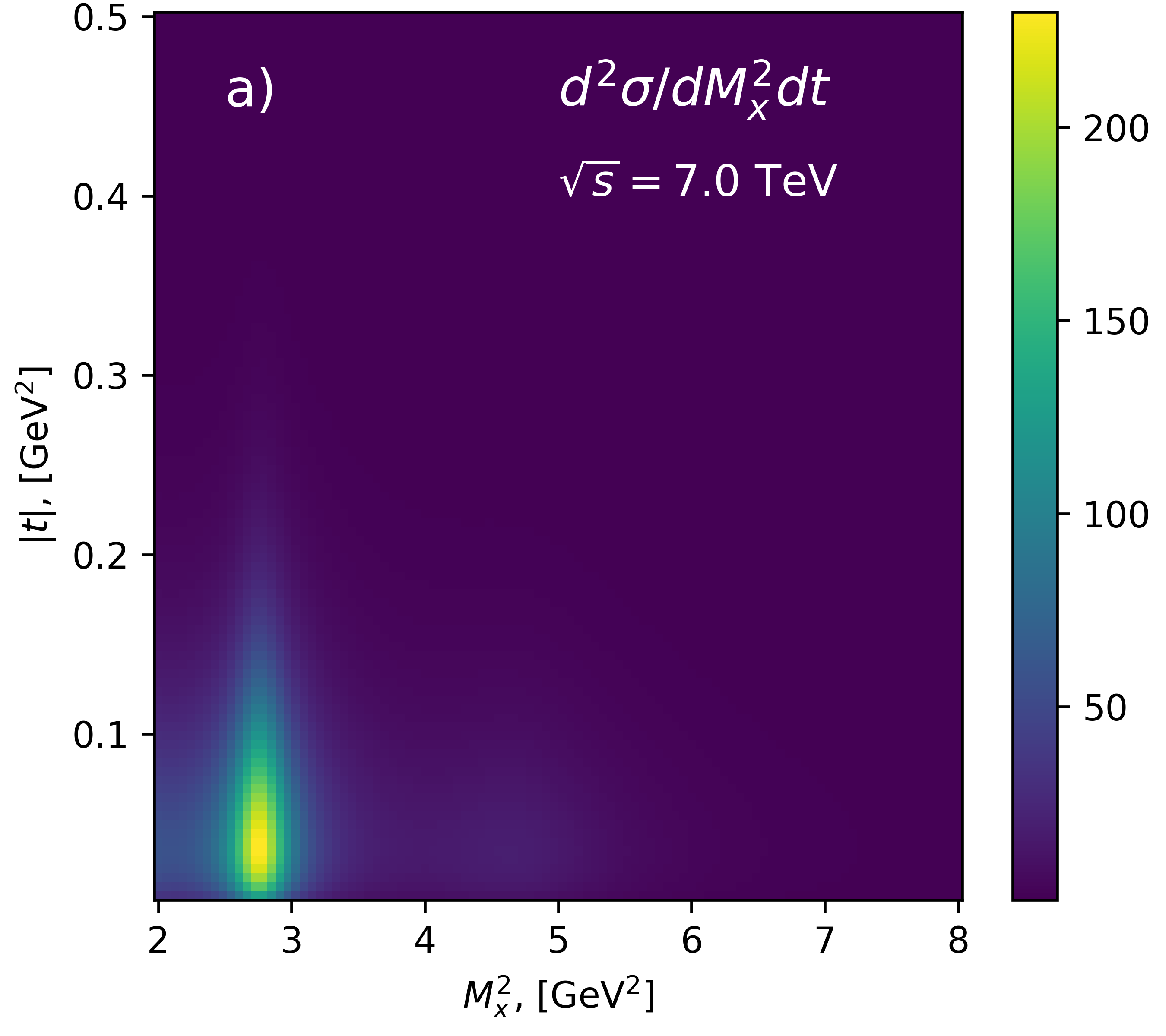}
\end{subfigure}%
\hfill
\begin{subfigure}{.49\textwidth}
  \centering
  \includegraphics[width=1\linewidth]{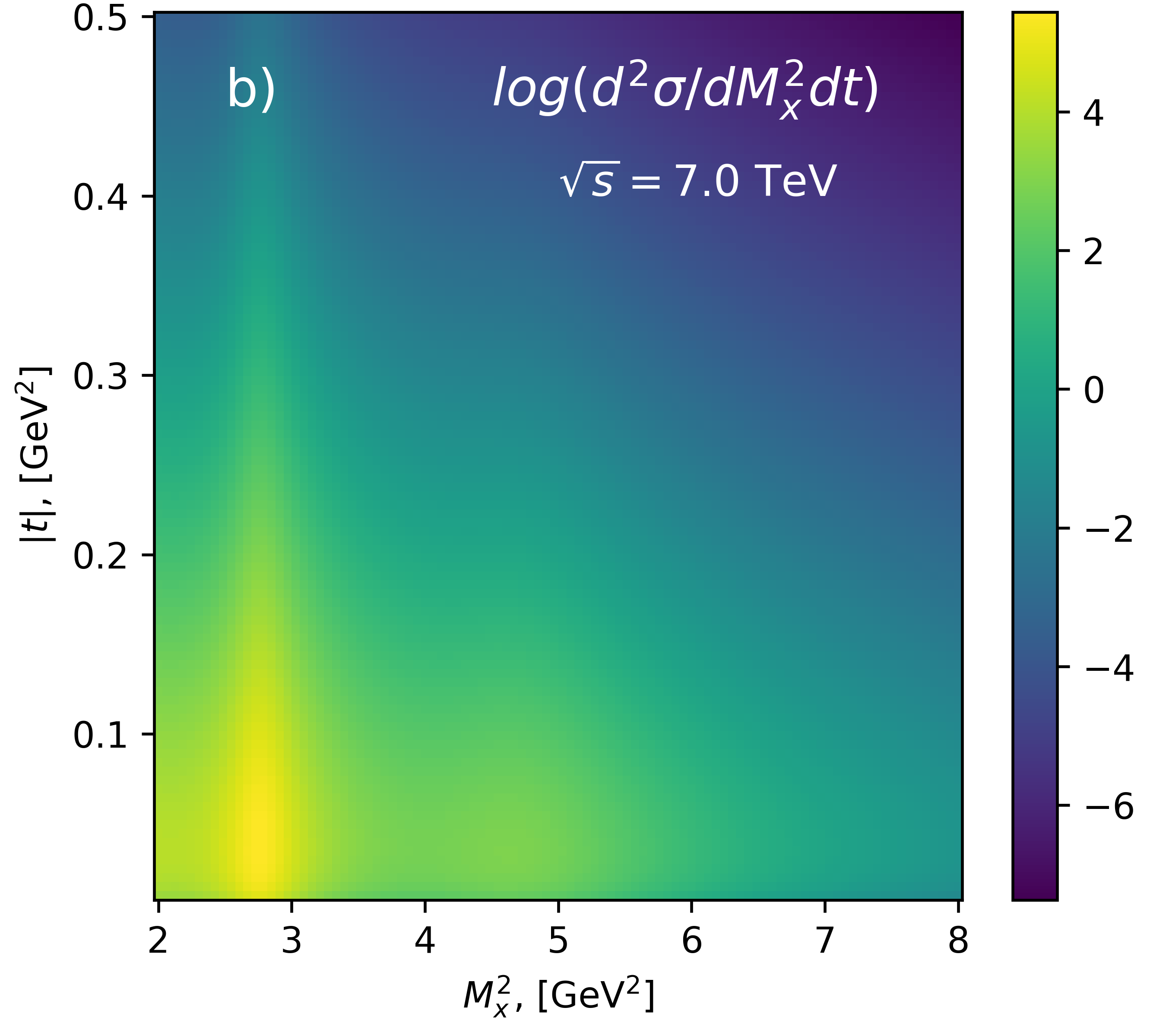}
\end{subfigure}
  \caption{The differential cross-section $d^2\sigma/dtdM_X^2$ \eqref{diffcs} in the $t$, $M_X^2$ plane (a), and its logarithm (b), plotted in the resonance region for $0 \leqslant -t \leqslant 0.5\;\text{GeV}^2$.}
\label{fig:diffcs}
\end{figure}

\begin{figure}[!ht]
\centering
\begin{subfigure}{.425\textwidth}
  \centering
  \includegraphics[width=1\linewidth]{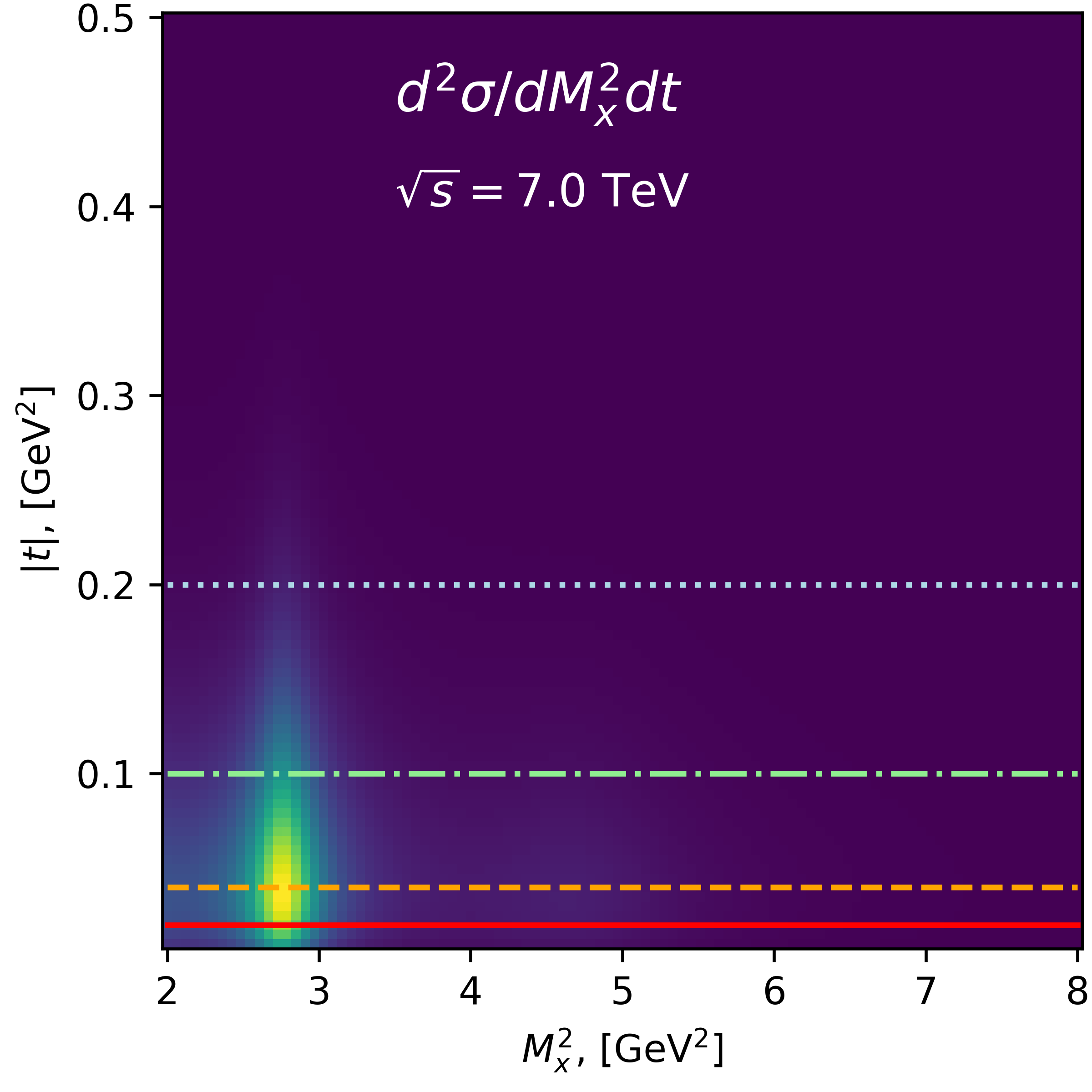}
\end{subfigure}%
\hfill
\begin{subfigure}{.55\textwidth}
  \centering
  \includegraphics[width=1\linewidth]{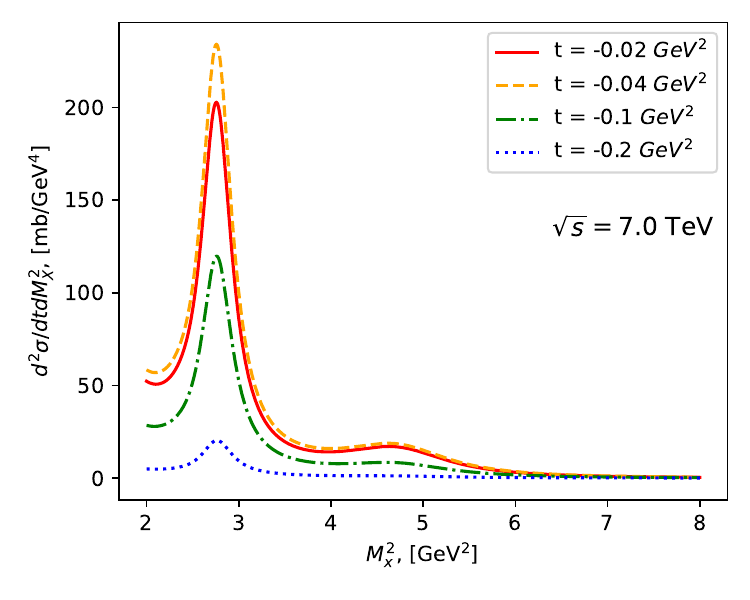}
\end{subfigure}
  \caption{The differential cross-section $d^2\sigma/dtdM_X^2$  \eqref{diffcs} in the $t$, $M_X^2$ plane in the resonance region for $0 \leqslant -t \leqslant 0.5\;\text{GeV}^2$ (left), and multiple regular plots in $M_X^2$ at fixed values of $t$ (right).}
\label{fig:diffcs_Mx2}
\end{figure}

\begin{figure}[!ht]
\centering
\begin{subfigure}{.425\textwidth}
  \centering
  \includegraphics[width=1\linewidth]{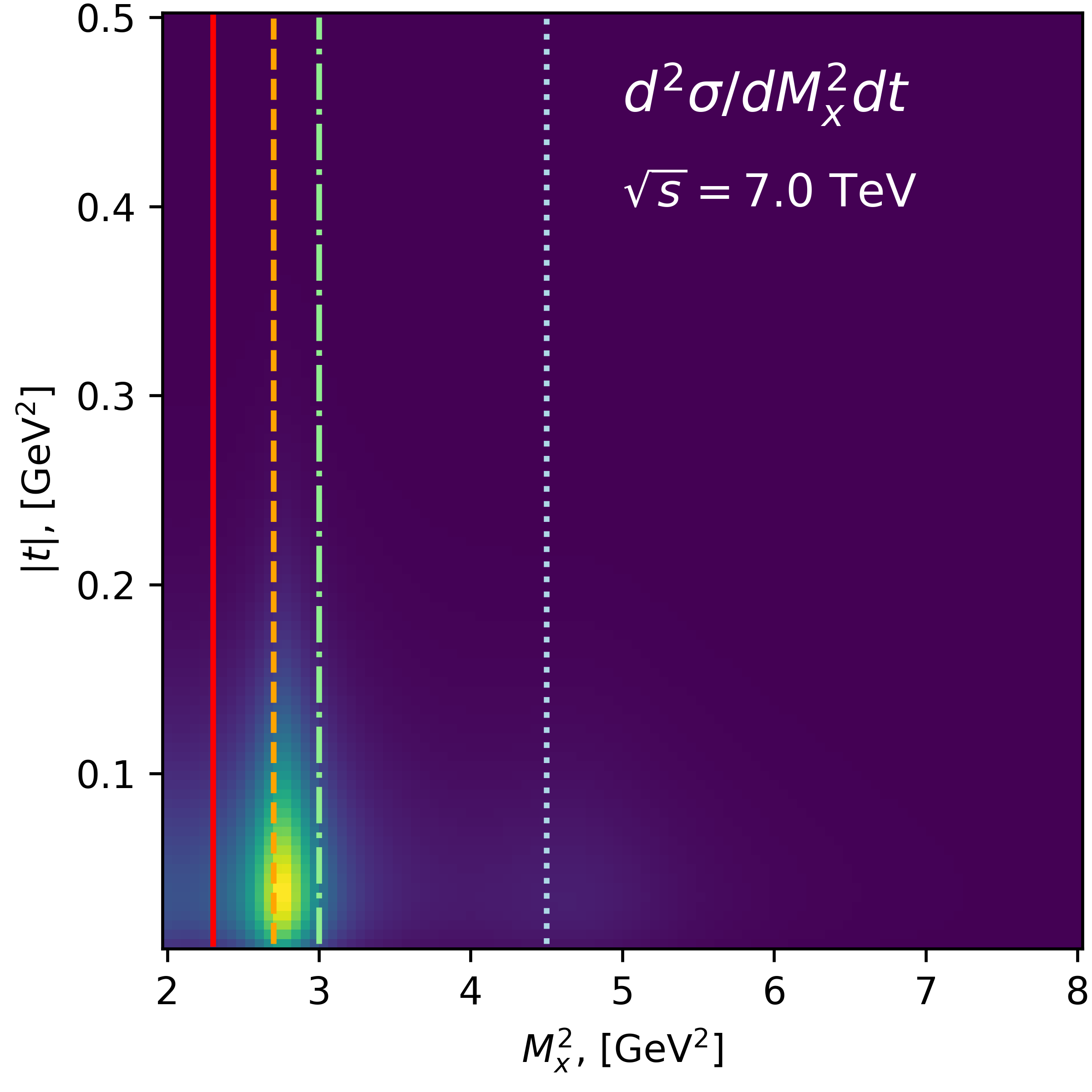}
\end{subfigure}%
\hfill
\begin{subfigure}{.55\textwidth}
  \centering
  \includegraphics[width=1\linewidth]{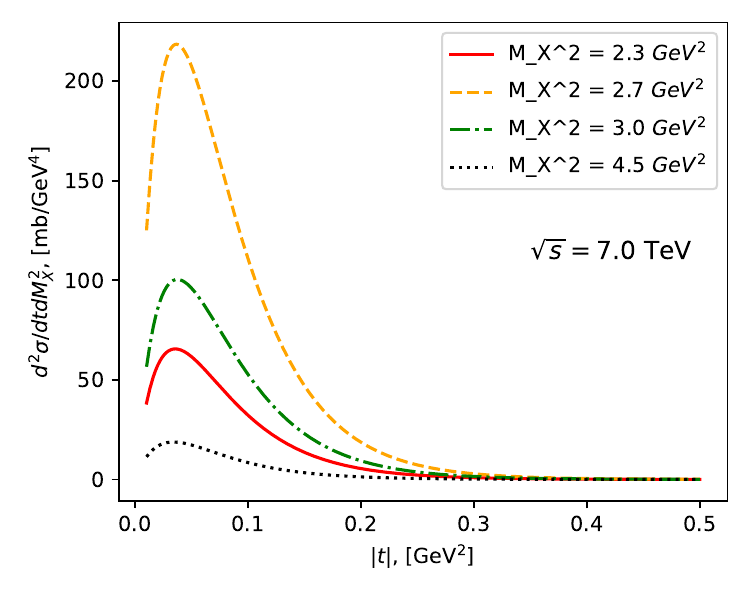}
\end{subfigure}
  \caption{The differential cross-section $d^2\sigma/dtdM_X^2$  \eqref{diffcs} in the $t$, $M_X^2$ plane in the resonance region for $0 \leqslant -t \leqslant 0.5\;\text{GeV}^2$ (left), and multiple regular plots in $t$ at fixed values of $M_X^2$ (right).}
\label{fig:diffcs_t}
\end{figure}

The plots of double differential cross-section \eqref{diffcs} in the resonance region are depicted on Figs.~\ref{fig:diffcs}, \ref{fig:diffcs_Mx2}, \ref{fig:diffcs_t}. 
The single peak in $M_X^2$ dimension is clearly visible in the resonance region.
The unfitted values of parameters used in this section are $A_0=10^3\;\text{mb/GeV}^2$ and $t_0 = 0.71\;\text{GeV}^2$.





\section{Cross-sections}
In this section we calculate total and differential cross-sections and fit them to the experimental data to find the values of parameters $A_0$ and $t_0$.

\begin{figure}[!ht]
\begin{center}
\includegraphics[width=0.75\linewidth]{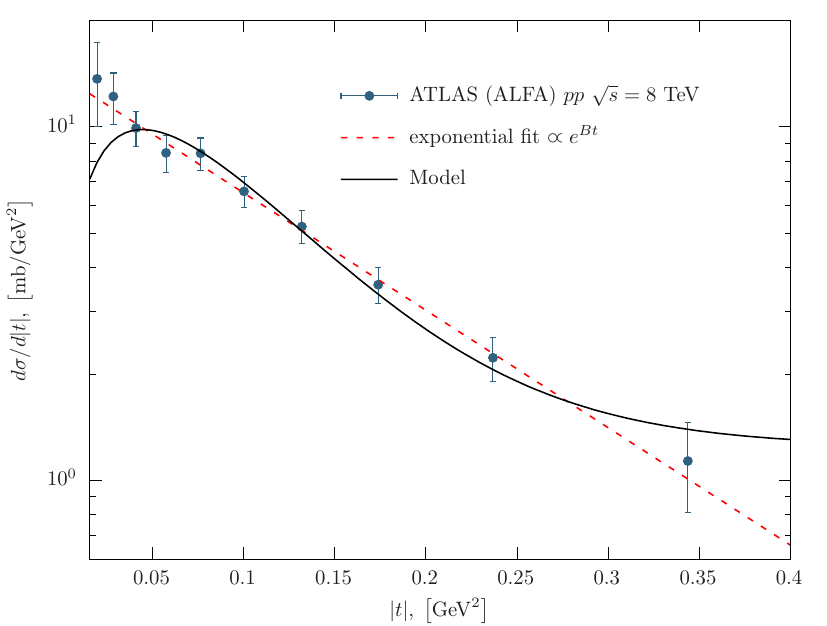}
\caption{
The differential cross-section $d\sigma/dt$ of single diffraction dissociation as a function of $|t|$.
The dashed line is the exponential fit to the experimental data \cite{Aad2020-cw}. 
The solid line is the model fit with the constant background contribution $b_0$.
}
\label{fig:difft}
\end{center}
\end{figure}

First, we integrate the double differential cross-section \eqref{diffcs} over $M_X^2$ in resonance region to calculate the differential cross-section

\begin{equation}
\frac{d\sigma}{dt}\left(t\right) = \int\limits_{2}^{8} \frac{d^2\sigma}{dtdM_X^2}\left(t, M_X^2\right)dM_X^2 + b_0,
\label{eq:difft}
\end{equation}

\noindent
where $b_0$ is the simple model of background contribution discussed in the previous section.
Then we fit \eqref{eq:difft} to the experimental data \cite{Aad2020-cw} using the \texttt{ROOT} implementation of \texttt{Minuit} framework. The fitting procedure converges with $\chi^2/d.o.f. \approx 1.07$, providing the following values of parameters: $A_0 = 35.58\;\text{mb/GeV}^2$, $t_0 = 1.486\;\text{GeV}^2$, and $b_0 = 8.2\;\text{mb/GeV}^2$. 
The result of fitting procedure is shown in \fref{fig:difft}.


\begin{figure}[!ht]
\begin{center}
\includegraphics[width=0.75\linewidth]{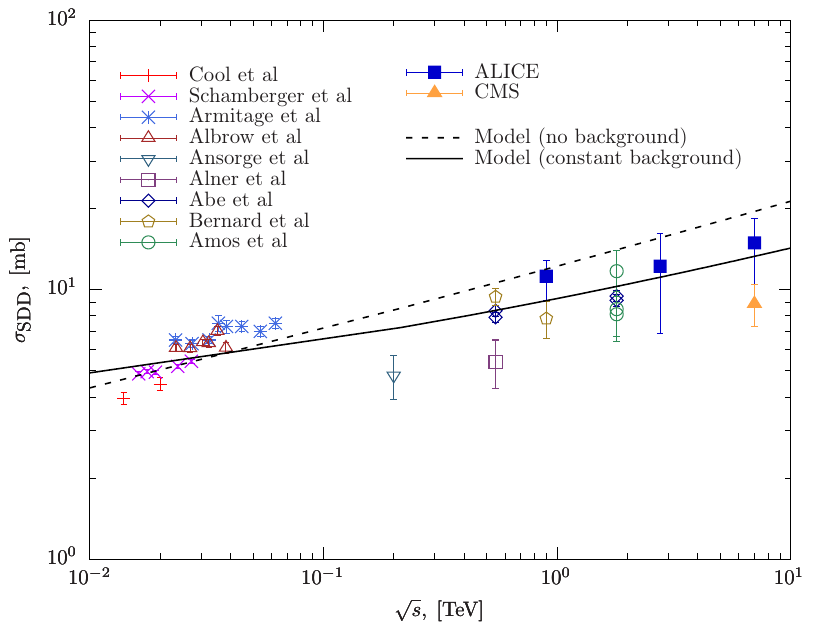}
\caption{
The total single diffraction dissociation cross-section as a function of $\sqrt{s}$. 
The dashed line is the model fit to the experimental data 
\cite{cs_cool, cs_schamberger, cs_armitage, cs_albrow, cs_ansorge, cs_alner, cs_abe, cs_bernard, cs_amos, cs_amos2, cs_khachatryan, cs_abelev} 
without any background contribution. 
The solid line is the fit with the constant background contribution $b$.
}
\label{fig:total}
\end{center}
\end{figure}

The next step, is to integrate \ref{eq:difft} over $t \in [0, 0.5]$ to calculate the total cross-section

\begin{equation}
\sigma_\text{SDD} = \int\limits_{-s}^{0} \int\limits_{2}^{8} \frac{d^2\sigma}{dtdM_X^2}\left(t, M_X^2\right)dM_X^2dt + b,
\label{eq:total}
\end{equation}

\noindent
where $b$ is the background contribution. 
Note, that we do not have any reliable model for the background, so we relax the connection between $b_0$ and $b$ and fit them separately.
The fit of differential cross-section $d\sigma/dt$ gives us the reasonable estimate for $t_0 = 1.486\;\text{GeV}^2$ that we can use in total cross-section fit.
The fitting procedure converges giving $\chi^2/d.o.f. = 14.03$ without background contribution, and providing he value of $A_0 = 565 \pm 3.11\;\text{mb/GeV}^2$. 
Accounting for the constant background contribution  $b$, gives us better fit result $\chi^2/d.o.f. = 10.72$. 
The values of parameters in this case are $A_0 = 378.43 \pm 16.68\;\text{mb/GeV}^2$ and $b = 1.85 \pm 0.16\;\text{mb/GeV}^2$.



\section{Low missing mass event generation}
The differential cross-sections fitted in the previous section provide facilities to generate events for low missing mass single diffractive processes. 

In this context, the four-momentum transfer squared $t$ and the squared mass $M_X^2$ of the dissociated system become random variables.
The joint probability density of $t$ and $M_X^2$ is given by the double differential cross-section \eqref{diffcs} at fixed $\sqrt{s}$

\begin{equation}
\rho\left(M_X^2, t\right) 
= 
\frac{1}{N}\frac{d^2\sigma}{dtdM_X^2}\left(M_X^2, t\right),
\label{density}
\end{equation}

\noindent
where $N = \sigma_{SDD}\left(\sqrt{s}\right)$ is the normalizing factor ensuring unit total probability, and $\sigma_{SDD}\left(\sqrt{s}\right)$ is the integrated single diffraction dissociation cross-section \eqref{eq:total}.
As discussed in the previous section, we consider the domain of $M_X^2 \in [2, 8]\;\text{GeV}^2$.

To generate an event we sample a pair $(t, M_X^2)$ from the probability density \eqref{density}.
While this can be done in multiple ways (e.g. the acceptance-rejection method \cite{Devroye1986}), we follow the algorithm used in the Minimum Bias Rockefeller (MBR) simulation \cite{CiesielskiMBR} implemented in Pythia 8 event generator.
The idea is to begin by sampling a value of $M_X^2$ from the marginal probability density function 
\begin{equation}
\rho_{M_X}\left(M_X^2\right) 
= 
\int\limits_{t_{min}}^{0}\rho\left(M_X^2, t\right) dt
=
\frac{1}{N}\int\limits_{t_{min}}^{0}\frac{d^2\sigma}{dtdM_X^2}\left(M_X^2, t\right)dt,
\label{Mx2density}
\end{equation}

\noindent
where $t_{min} = -1\;\text{GeV}^2$ defines the integration lower bound, cutting  off the region of the small values of cross-section.
The integral in \eqref{Mx2density} is calculated numerically.

Next, the value of $t$ is sampled from the conditional probability density function
\begin{equation}
\rho_{t}\left(t\;\middle|\;M_X^2\right) 
= 
\frac{\rho\left(M_X^2, t\right) 
}{\rho_{M_X}\left(M_X^2\right) 
}.
\label{density}
\end{equation}

\noindent
Finally, the generated pair $(t, M_X^2)$ together with the Mandelstam variable $s$ gives us the four-momenta of the proton and disscociated system in the final state, which can be calculated from the relativistic kinematics of two-body scattering.

\begin{figure}[!ht]
\begin{center}
\includegraphics[width=0.6\linewidth]{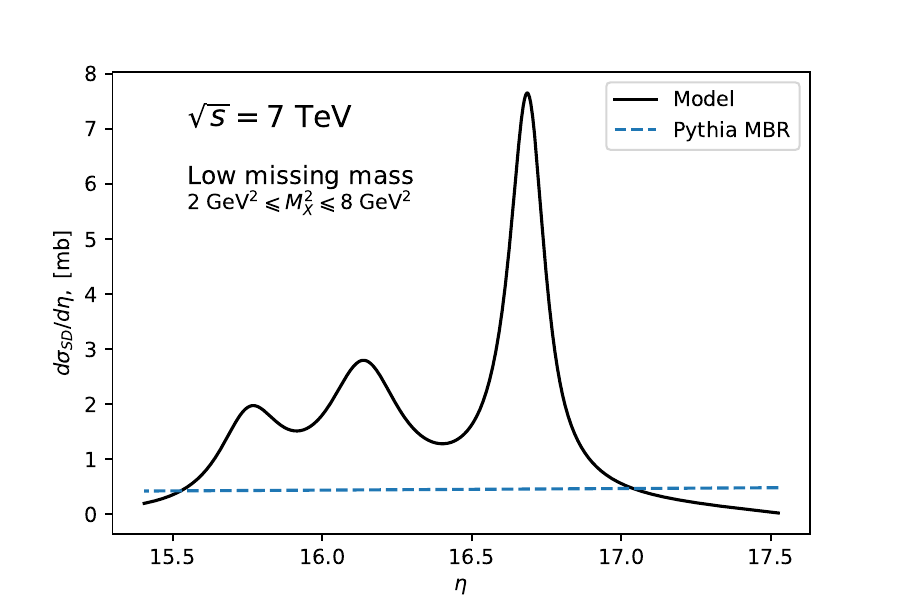}
\caption{The comparison of the our model agains the MBR simulation in Pythia 8 at $\sqrt{s} = 7$ TeV.}
\label{fig:mbr_vs_jenk}
\end{center}
\end{figure}

To compare our results against the MBR approach, alongside with kinematic variable $M_X^2$ we consider the rapidity gap $\eta = -\log\left(M_X^2/s\right)$. 
In this variable, the low missing mass region $M_X^2 \in [2, 8]\;\text{GeV}^2$ corresponds to the interval  $15.6 \leqslant \eta \leqslant 17$ at $\sqrt{s} = 7\;\text{TeV}$.
Despite this interval is quite narrow compared to the $0 \leqslant \eta \leqslant 17.5$ in \cite{CiesielskiMBR},
it exhibits a significant difference between two approaches.
The MBR simulation demonstrates the close to constant behavior of the cross-section in the region of low missing masses.
Whereas our model predicts highly non-monotonical dependency with multiple peaks corresponding to the resonances.


\section{Summary}
The in-depth examination of the differential cross-sections behavior within the resonance region is performed, particularly at low missing masses $M_X$. 
The refinement of model parameters through the incorporation of new experimental data is discussed, improving accuracy and predictive capabilities.
The fits of differential and total cross-sections were performed using C++ programs within the ROOT framework, specifically employing Minuit for parameter optimization. Results are presented as numerical values of parameters, goodness-of-fit metrics, and graphical representations, providing a clear and comprehensive depiction of the model performance.
The constant background contribution $b$ was assumed due to the lack of a refined model for background effects. 
The proper modeling of background contributions could potentially yield significant improvements in the accuracy of results and should be a focus of future research.

\section*{Aknowledgments}
This project has received funding through the EURIZON project, which is funded by the European Union under grant agreement No.871072.

\bibliographystyle{unsrtnat}
\bibliography{references}

\end{document}